# Self-diffusion in carbon-alloyed CoCrFeMnNi high entropy alloys

O.A. Lukianova[1,*], V. Kulitckii[1], Z. Rao[2], Z. Li[2,3], G. Wilde[1], S.V. Divinski[1,**]

[1]Institute of Materials Physics, University of Münster, Germany
[2]Max-Planck-Institut für Eisenforschung, Düsseldorf, Germany
[3]School of Materials Science and Engineering, Central South University, Changsha, China

*Corresponding authors: * sokos100@mail.ru  ** divin@uni-muenster.de

Tracer diffusion of the substitutional components in $(CoCrFeNiMn)_{1-x}C_x$ high-entropy alloys with $x = 0.002$, $0.005$ and $0.008$ (in at. fractions) is measured at elevated temperatures from 1173 to 1373 K. Two different characteristic effects of interstitial carbon addition on substitutional diffusion in these FCC alloys are distinguished. At the highest temperature of 1373 K, alloying by C with relatively low concentrations ($x = 0.002$) retards diffusion of the substitutional elements with respect to those in the C-free alloy. At lower temperatures and/or higher C concentrations ($x \geq 0.005$), an enhancement of the diffusion rates of all substitutional elements is seen. A model is suggested that relates the self-diffusivities in the CoCrFeMnNi-C alloys with the lattice distortion imposed by interstitially dissolved carbon. The experimental results are interpreted in terms of a decrease of the migration barriers for vacancy-mediated diffusion due to the presence of interstitial C atoms.

## 1. Introduction

Nowadays multi-principal element alloys, (also called termed as high entropy alloys (HEAs) [1–4], have undoubtedly attracted a broad interest in Materials Science. Along with many attractive properties, however, one of the most noticeable disadvantages of single-phase face centered cubic (FCC) structured HEAs is their relatively low strength at room temperature, only about 200 MPa in the as-cast state [1] that is significantly less than the intended strength of metallic structural materials. Grain refinement and precipitation strengthening, being quite effective for improving the strength of HEAs [5–9], may lead to a loss of plasticity. Carbon alloying is recognized as an effective method of increasing the strength of HEAs while at the same time without sacrificing ductility [7, 9].

The equiatomic CoCrFeMnNi HEA with FCC crystalline structure is one of the most promising alloys because of its beneficial combination of mechanical and physical properties [10]. Recent experimental studies have clearly demonstrated that its alloying by interstitial carbon (and/or nitrogen) improves both strength and plasticity [5, 10, 11–18]. This behavior results from co-influence of Twinning Induced Plasticity (TWIP), precipitation, and solid solution hardening mechanisms [10, 19, 20]. Peng et al. [17] discussed in detail the carbide precipitation as a hardening mechanism in the CoCrFeMnNi-based interstitial alloys (in addition to, e.g., solid solution hardening).

Significant distortions of the crystal lattice of a high-entropy alloy, considered as one of the so-called core effects [1], contribute to the appearance of additional stresses and might even lead to the improvement of mechanical properties. While solid solution hardening in HEAs attracted a considerable interest [21–24], the interstitial solid solution hardening has not received much attention and has not yet been studied in detail.

Being important for mechanical properties, the interstitial elements might affect the diffusion jumps of substitutional atoms as it was found in the case of austenite [25] or suggested for copper [26]. Since diffusion plays an important role in a number of high-temperature properties of structural materials or even determines some of them [10], reliable data on diffusion of substitutional elements are essential for the development of new materials and for the evaluation of their long-term stability.

In our previous work [27], we measured the tracer diffusion coefficients in the CoCrFeMnNi–C alloys at 1373 K. The present work is focused on a detailed study of substitutional diffusion in the FCC CoCrFeMnNi-based HEAs with various carbon contents in an extended temperature interval and evaluation of the carbon effect on the crystal lattice parameter in these alloys. The available tracer diffusion coefficients for all constituting elements in C-free CoCrFeMnNi [27–31] are used as a reference. The impact of interstitial C on the elastic distortions in the HEAs is evaluated, too.

## 2. Experimental details
### 2.1 Materials preparation

The equiatomic CoCrFeMnNi HEAs with different carbon contents were synthesized using technically pure metallic elements and carbon by melting and casting using a vacuum induction furnace. The details of the thermo-mechanical processing and measured bulk chemical compositions of the as-cast alloys as determined by wet-chemical analysis are given in Ref. [10]. The materials in the present study are the same as produced and used in Ref. [10].

The alloy sheets were first hot-rolled at 1223 K to 50 % reduction in thickness and subsequently homogenized at 1473 K for 3 h in Ar atmosphere followed by water-quenching. This processing maximized the amount of dissolved carbon in the FCC matrix.

**2.2 Microstructure characterization**

X-ray diffraction measurements were performed by a X6_WS diffractometer with Cu K radiation operating at 40 kV and 30 mA in θ–2θ geometry between 30° and 130° with a step size of 0.05° and a counting time of 20 s per step.

The microstructure has been analyzed on samples used later for the diffusion measurements. Disc-shaped samples of 8 mm in diameter and about 1.5 mm in height were cut by spark-erosion. Subsequently, one face was polished to mirror-like finish using standard metallography procedures. The samples were again annealed in purified Ar atmosphere at 1373 K for 24 h to remove the induced mechanical stresses. After annealing treatment, the samples were fine polished using an oxide suspension (OPS) with silica particle sizes around 50 nm for more than 30 min. Finally, the samples were polished with soap and ethanol for 5 min to remove the nanosilica particles. Some samples were further annealed to mimic the annealing treatments for diffusion measurements (at 1173 K for 2 days and 12 hours) and characterize the microstructure evolution during diffusion.

Electron back-scatter diffraction (EBSD) measurements were carried out on a Zeiss Crossbeam XB 1540 FIB scanning electron microscope (SEM) with a Hikari camera and the TSL OIM software. Back-scattered electron (BSE) and secondary electron (SE) images were taken on a Zeiss-Merlin instrument, and energy-dispersive X-ray spectroscopy (EDS) was used to identify the elemental distributions on a micrometer scale.

**2.3 Radiotracer experiments**

A standard radiotracer technique [32] was used to determine diffusion coefficients of all substitutional elements, i.e. of Co, Cr, Fe, Mn, and Ni. The radiotracer solutions (either a single $Ni^{63}$ β-isotope or a mixture of $Co^{57}$, $Cr^{51}$, $Fe^{59}$, and $Mn^{54}$ γ-isotopes) were applied on a polished sample surface and dried under an infrared lamp. The samples were packed in a tantalum foil, sealed into silica glass tubes under a purified Ar atmosphere, and diffusion annealed at selected temperatures for the given times. To exclude any interference with lateral and/or surface diffusion, the samples were subsequently reduced in diameter by at least 1 mm.

Table 1 Measured densities of the studied alloys.

| Carbon content, at. % | 0.2 | 0.5 | 0.8 |
|---|---|---|---|
| Density, g/cm$^3$ | 8.002±0.001 | 7.987±0.001 | 7.975±0.001 |

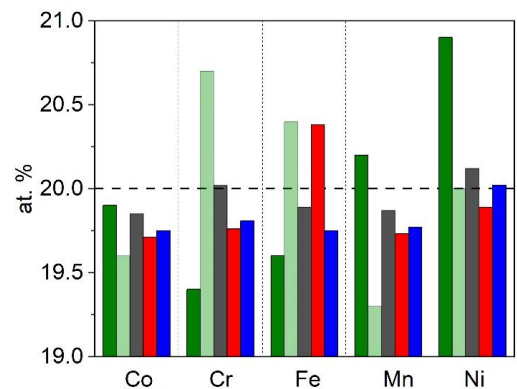

Fig. 1. Relative composition (normalized on the total amount of substitutional elements) in the present and reference [28, 29, 31] HEAs. Green, black, red and blue bars present the chemical compositions of alloys with 0, 0.2, 0.5 and 0.8 at.% of carbon, respectively. The composition of the C-free CoCrFeMnNi in the presented work and in Refs. [28, 31] are quite close (dark green bars). The composition of C-free CoCrFeMnNi from Ref. [29] is highlighted by light green.

The penetration profiles were determined by parallel mechanical sectioning using a precision grinding machine and a Mylar foil with 30 μm large SiC particles. The section masses were determined by weighing the samples before and after each sectioning step on a microbalance to an accuracy of ±0.1 μg. The specific activity of each section was measured by a pure Ge γ-detector equipped with a 16 K multi-channel analyzer (the $Co^{57}$, $Cr^{51}$, $Fe^{59}$, and $Mn^{54}$ radioisotopes). The energies of the corresponding γ-decays can easily be distinguished by the available setup. The activity of the $Ni^{63}$ radioisotope (β-decays) was analyzed by a Liquid Scintillation Counter TRI-CARB 2910. The background corrected specific radioactivity divided by the section mass is proportional to the tracer concentration in the given section.

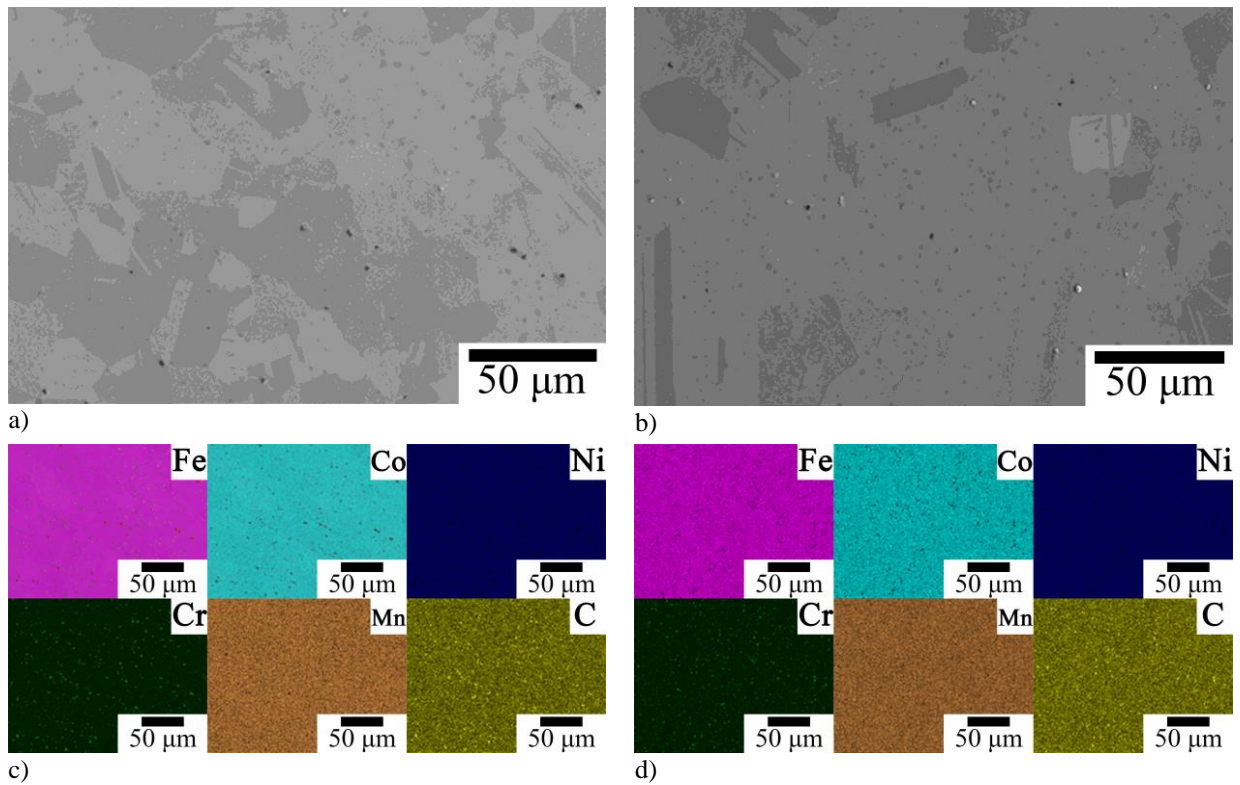

Fig. 2. Microstructure and elemental distributions of the CoCrFeMnNi–C HEAs with 0.2at. % C after annealing at 1373 K for 12 hours. (a, c) and 48 hours (b, d) BSE images (a, b) and EDS maps (c, d) show the morphologies of the microstructures and the distributions of the constituent elements, respectively.

The density of the reported alloys was measured by weighing the samples using the Archimedes' method, see Table 1, and used for the reliable determination of the penetration depths.

## 3. Results

### 3.1 Chemical composition

Since changes of local chemical environments could affect vacancy-mediated diffusion [31], we compared the nominally equiatomic compositions of the studied HEAs with respect to the relative amounts of 3d transition metallic elements, Fig. 1.

In all alloys, the substitutional elements are kept at almost equiatomic proportions. Polycrystalline C-free HEA studied in Ref. [29] is seen to be slightly enriched in Cr and depleted with Mn. The present C-free CoCrFeMnNi alloy (identical to that investigated by Gaertner et al. [28, 31]) reveals almost an opposite trend with respect to these elements. The C-bearing alloys show almost uniform distributions of the substitutional elements, Fig. 1.

### 3.2 Microstructure characterization

As an example, the microstructures of the reported alloys with 0.2 at. % of carbon annealed at 12 and 48 hours are shown in Fig. 2. Twins and carbides are clearly seen in the studied materials. It was proven that the microstructure remains practically unchanged after annealing treatments mimicking the diffusion annealing. As it was described in [10] the interstitial HEA with a nominal carbon content of 0.2 at. % is characterized by a fully recrystallized microstructure with a large fraction of annealing twins. The average grain size is approximately 45 µm. Nano-particles with an average size of 50 to 100 nm and enriched in Cr are found to be randomly distributed in the microstructure. It has been shown that the annealing time does not affect the microstructure. Such microstructure is suitable for reliable volume diffusion measurements.

### 3.3 Impact of C alloying on lattice constant

In Fig. 3a, a FCC unit cell of the Cantor alloy (the elements are shown as coloured spheres of different radii to mimic the CoCrFeMnNi alloy) is sketched with an additional C atom (grey sphere) at an octahedral position. The actual distribution of constituent atoms over the lattice sites is assumed to be random. Carbon alloying increases the lattice constant. In Fig. 3b, the results of the present XRD measurements (open circles) are compared with the available literature data (squares and triangles).

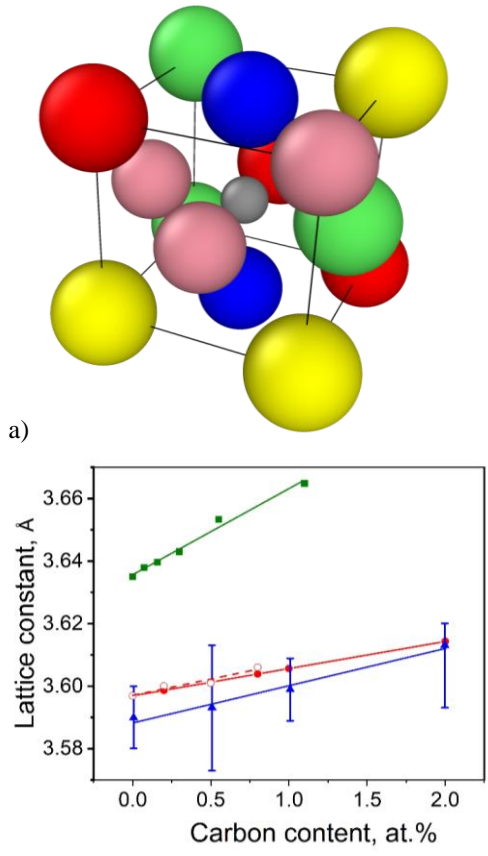

Fig. 3 a) FCC unit cell of a five-component HEA (coloured spheres) with an interstitial carbon at an octahedral position (grey sphere). b) The measured lattice constant (open red circles) as function of the carbon content in comparison to the literature data (green squares [33] and blue triangles [34]) and the predictions according to Eq. (1) (filled red circles).

### 3.3.1. C-free Cantor alloy

In the present work, the lattice constant of equiatomic CoCrFeMnNi alloy was measured at 0.3589 nm. If Vegard's law would be used, a slightly smaller value of 0.3585 nm is obtained. The Lubarda approach [35] proposed by Moreen [36] and Toda-Caraballo [37] predicts the lattice constant as 0.3597 nm. These results agree reasonably well with the present data and the reported values of the crystal lattice parameter of equatomic CoCrFeMnNi: 0.3595 nm [38], 0.359 nm [39], 0.3592 nm [40], and 0.3577 nm [41]. A lattice constant of 0.34685 nm was reported for the high-entropy chromium-depleted alloy [42]. For Fe-rich non-equatomic CoCrFeMnNi HEA ($Fe_{40}Mn_{28}Ni_8Cr_{24}$), values of 0.3615 nm [34] and 0.362 nm [43] were found.

### 3.3.2 The CoCrFeMnNi–C alloys

The Vegard approximation cannot be used for the case of interstitial alloying and we followed the approach suggested for austenite [44]. Rammo et al. [45] have shown that the lattice constant of austenite, $a$, increases due to carbon alloying following a semi-empirical formula,

$$a = a_0 + \frac{2(r_{Fe}+r_C)-2\sqrt{2}r_{Fe}}{4\cdot 10}X_C, \quad (1)$$

where $X_C$ is the concentration of C atoms in at.%, $r_{Fe}$ and $r_C$ are the atomic radii of Fe and C atoms. This equation was used with an averaged value of atomic radii of the matrix atoms in the CoCrFeMnNi alloy, $\bar{r}$, and $\bar{r}$ was determined by the Lubarda approach [35] for the C-free CoCrFeMnNi alloy. Equation (1) predicts a linear increase of the lattice constant of C-alloyed CoCrFeMnNi and a good agreement with the experiment is seen, Fig. 2b.

### 3.4 Tracer diffusion

As an example, the penetration profiles for diffusion of substitutional elements in the CoCrFeMnNi alloys with 0, 0.2, 0.5 and 0.8 at.% of C measured at 1273 K are shown in Fig. 4 in the coordinates of the logarithm of concentration vs. the depth squared, $y^2$. Two branches, identified with volume and grain boundary diffusion contributions, are recognized. In the present work, we are focusing on reliable measurements of volume diffusion for all isotopes and we will not report the impact of C alloying on the grain boundary diffusion rates of the substitutional elements. The penetration profiles were fitted accounting for both, volume and grain boundary diffusion contributions, solid lines in Fig. 4, and true volume diffusion data were extracted.

The lowest value of the volume diffusion coefficient, being inversely proportional to the slope in the given coordinates, was observed for the C-free CoCrFeMnNi alloy. The alloy with 0.8 at. % of C has obviously the highest value of the volume diffusion coefficient. The same trend is observed for other temperatures.

The instantaneous source solution of the diffusion problem has been found to be appropriate and the determined tracer diffusion coefficients are listed in Table 2 and are plotted in Fig. S1 (Supplementary Material) in Arrhenius coordinates.

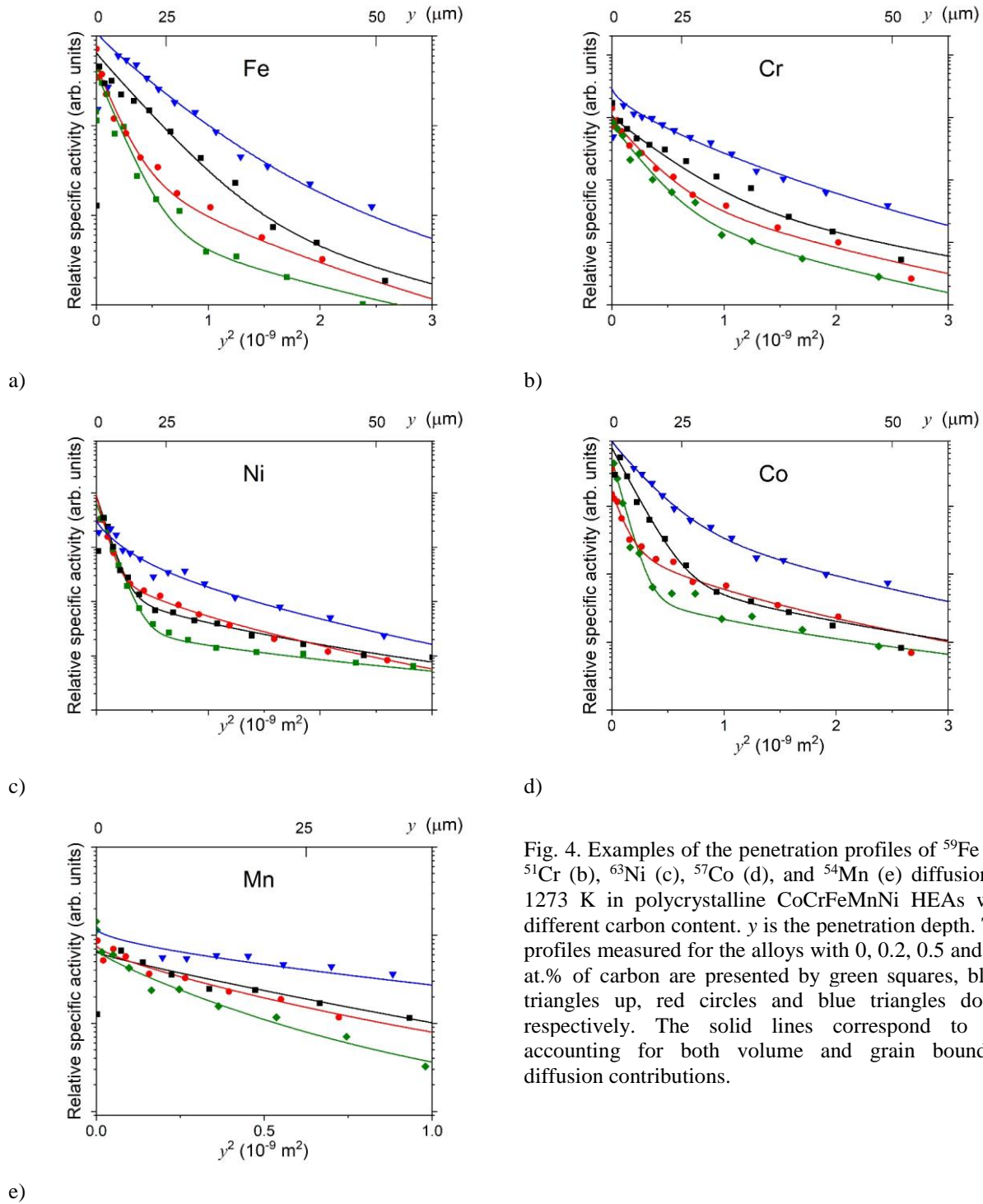

Fig. 4. Examples of the penetration profiles of $^{59}$Fe (a), $^{51}$Cr (b), $^{63}$Ni (c), $^{57}$Co (d), and $^{54}$Mn (e) diffusion at 1273 K in polycrystalline CoCrFeMnNi HEAs with different carbon content. $y$ is the penetration depth. The profiles measured for the alloys with 0, 0.2, 0.5 and 0.8 at.% of carbon are presented by green squares, black triangles up, red circles and blue triangles down, respectively. The solid lines correspond to fits accounting for both volume and grain boundary diffusion contributions.

In Fig. 5, the measured diffusion coefficients of all substitutional elements are shown as functions of both inverse temperature and the carbon content (the filled spheres). The separate plots are given in Supplementary Material, Figs. S1 and S2. The results for C-alloyed CoCrFeMnNi substantiate the trend observed previously for C-free CoCrFeMnNi HEA [28–30]. Indeed, Mn is the fastest element and Co and Ni are slowest elements in the temperature interval of 1173 to 1373 K. Cr and Fe reveal similar fast diffusivities in all alloys, Fig. 5.

## 4. Discussion

Presently, an extensive information on volume diffusion of all constituent elements in the five-component CoCrFeMnNi HEA is accumulated [27,29–31,51,52]. The reported values are generally in very good agreement, although slight variations of the composition (nominally equiatomic) need to be considered. Carbon could be present in HEAs as a spurious element due to manufacturing conditions. Since the diffusion properties are of key importance for predicting the life-time properties and the phase stability, knowing the impact of interstitial carbon and carbide formation on the atomic transport is imperative for developing HEA for advanced applications. Below we will focus on the impact of dissolved carbon on tracer diffusion of substitutional atoms in this alloy system.

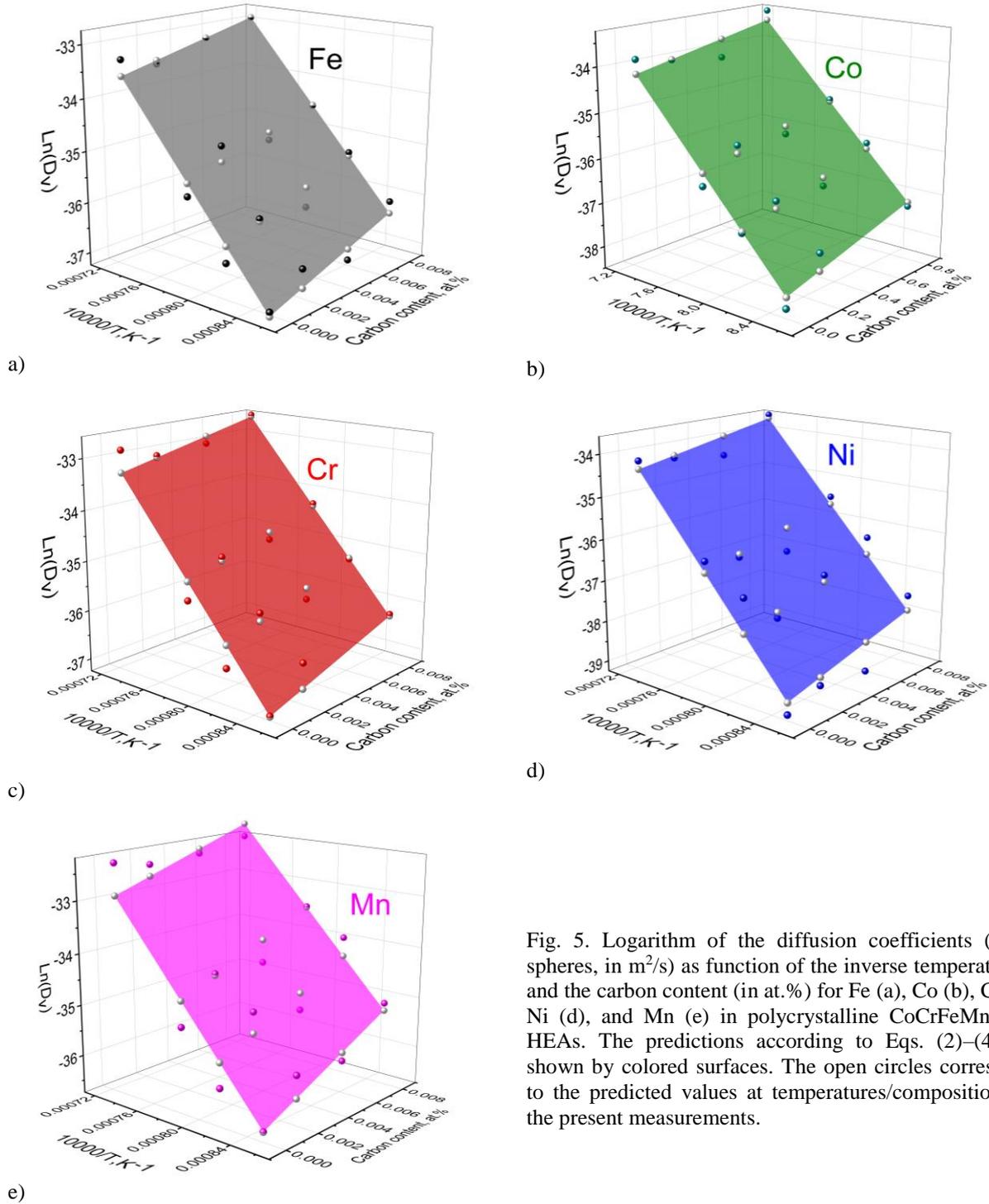

Fig. 5. Logarithm of the diffusion coefficients (filled spheres, in m$^2$/s) as function of the inverse temperature $T$ and the carbon content (in at.%) for Fe (a), Co (b), Cr (c), Ni (d), and Mn (e) in polycrystalline CoCrFeMnNi–C HEAs. The predictions according to Eqs. (2)–(4) are shown by colored surfaces. The open circles correspond to the predicted values at temperatures/compositions of the present measurements.

### 4.1 Tracer diffusion

Mead and Birchenall [25] analysed self-diffusion of iron in austenite with different carbon contents. Using the model of Overhauser [26], a linear increase of the logarithm of the tracer diffusion coefficient of Fe with C concentration has been predicted. Adopting this model, following relation is used to describe substitutional diffusion in the CoCrFeMnNi–C alloys,

$$D(x_C) = D(0)exp[bx_C]$$
$$= D(0)exp\left[A\left(\frac{B\Omega_m}{k_B T}\right)^2 x_C\right]. \quad (2)$$

Here $D(x_C)$ is the diffusion coefficient of the $(CoCrFeMnNi)_{1-x_C}C_{x_C}$ alloy, $D(0)$ is the diffusion coefficient in the C-free alloy, $x_C$ is the carbon concentration (in at. fractions), $A$ is a numerical factor, $B$ is the bulk modulus, $\Omega_m$ is the migration volume of vacancy-mediated diffusion, and $k_B T$ has its usual meaning. In the original work [26], the constant $b$ in Eq. (2) was estimated as,

$$b \cong \frac{128\pi}{45}\left(\frac{gK}{k_B T}\right)^2 \left(\frac{a}{R_0}\right)^3. \quad (3)$$

Table 2. Tracer diffusion coefficients in polycrystalline CoCrFeMnNi–C HEAs.

| T, K | $D$ ($10^{-16}$ m$^2$/s) | | | | |
|---|---|---|---|---|---|
| | Co$^{57}$ | Cr$^{51}$ | Fe$^{59}$ | Mn$^{54}$ | Ni$^{63}$ |
| C = 0 at.% | | | | | |
| 1373 | $20.1^{+0.1}_{-0.1}$ | $54.6^{+0.1}_{-0.1}$ | $35.4^{+0.2}_{-0.2}$ | $94.4^{+0.2}_{-0.2}$ | $14.7^{+0.3}_{-0.1}$ |
| 1273 | $1.84^{+0.01}_{-0.03}$ | $4.60^{+0.04}_{-0.26}$ | $3.69^{+0.01}_{-0.19}$ | $5.80^{+0.19}_{-0.02}$ | $2.01^{+0.01}_{-0.02}$ |
| 1223 | $0.862^{+0.103}_{-0.005}$ | $1.37^{+0.09}_{-0.05}$ | $1.32^{+0.21}_{-0.17}$ | $2.28^{+0.02}_{-0.57}$ | $1.10^{+0.01}_{-0.01}$ |
| 1173 | $0.234^{+0.003}_{-0.014}$ | $0.73^{+0.029}_{-0.046}$ | $0.70^{+0.030}_{-0.028}$ | $1.33^{+0.17}_{-0.05}$ | $0.10^{+0.039}_{-0.022}$ |
| C = 0.2 at.% | | | | | |
| 1373 | $17.1^{+0.3}_{-0.3}$ | $43.3^{+1.1}_{-0.1}$ | $28.5^{+0.2}_{-0.2}$ | $81.5^{+0.2}_{-0.4}$ | $13.4^{+0.3}_{-0.3}$ |
| 1273 | $3.53^{+0.10}_{-0.07}$ | $7.65^{+0.79}_{-1.03}$ | $7.78^{+0.49}_{-0.12}$ | $1.30^{+0.04}_{-0.03}$ | $1.73^{+0.07}_{-0.04}$ |
| 1223 | $1.30^{+0.24}_{-0.01}$ | $3.09^{+0.14}_{-0.03}$ | $2.36^{+0.45}_{-0.04}$ | $7.53^{+0.02}_{-0.12}$ | $0.51^{+0.009}_{-0.005}$ |
| 1173 | $0.56^{+0.50}_{-0.050}$ | $0.15^{+0.006}_{-0.007}$ | $1.17^{+0.42}_{-0.02}$ | $2.87^{+0.02}_{-0.07}$ | $0.14^{+0.002}_{-0.005}$ |
| C = 0.5 at.% | | | | | |
| 1373 | $14.4^{+0.2}_{-0.2}$ | $45.7^{+1.5}_{-0.7}$ | $39.3^{+1.1}_{-0.1}$ | $85.1^{+0.1}_{-0.2}$ | $11.3^{+0.6}_{-0.2}$ |
| 1273 | $3.28^{+0.15}_{-0.30}$ | $8.12^{+0.12}_{-0.59}$ | $6.57^{+0.22}_{-0.40}$ | $12.2^{+2.10}_{-12.7}$ | $1.39^{+0.05}_{-0.03}$ |
| 1223 | $1.23^{+0.01}_{-0.80}$ | $2.86^{+0.25}_{-0.04}$ | $2.07^{+0.08}_{-0.05}$ | $5.63^{+0.21}_{-0.15}$ | $0.69^{+0.009}_{-0.006}$ |
| 1173 | - | - | $0.92^{+0.011}_{-0.078}$ | $2.56^{+0.03}_{-0.03}$ | $0.12^{+0.002}_{-0.001}$ |
| C = 0.8 at.% | | | | | |
| 1373 | $35.2^{+0.2}_{-0.1}$ | $70.4^{+0.1}_{-0.2}$ | $50.3^{+0.7}_{-1.7}$ | $10.3^{+0.1}_{-0.2}$ | $25.6^{+0.3}_{-0.2}$ |
| 1273 | $5.44^{+0.14}_{-0.30}$ | $13.2^{+0.20}_{-0.50}$ | $10.1^{+1.00}_{-2.00}$ | $2.94^{+0.11}_{-0.16}$ | $3.99^{+0.16}_{-0.09}$ |
| 1223 | $2.32^{+0.54}_{-0.09}$ | $4.75^{+0.14}_{-0.45}$ | $4.53^{+0.16}_{-0.13}$ | $1.79^{+0.03}_{-0.23}$ | $1.69^{+0.04}_{-0.11}$ |
| 1173 | $0.66^{+0.028}_{-0.003}$ | $1.83^{+0.18}_{-0.39}$ | $2.03^{+0.82}_{-0.81}$ | $5.70^{+1.12}_{-0.10}$ | $0.49^{+0.029}_{-0.021}$ |

Here $g$ is a dimensionless constant which determines the displacements of substitutional atoms from their ideal lattice positions due to incorporation of an interstitial C atom as a spherical elastic singularity at the octahedral position and its value for FCC Cu has been estimated at 0.06 [26]; $K = \frac{-\Delta E_m}{\epsilon}$ is the change of the barrier height at the saddle position, $\Delta E_m$, due to the elastic strain $\epsilon$ induced by an interstitial carbon atom; $R_0$ is of the order of the lattice constant $a$ and corresponds to the cutoff of the elastic distortions around C atom.

Re-writing Eq. (1) with respect to the elastic distortions induced by carbon atoms, $\delta_C$, $\delta_C = \frac{a-a_0}{a_0}$, and expressing the C concentration in at. fractions, $x_C$, one may correlate the C-induced changes of the diffusion coefficients, $D(x_C)$, with the local distortions and one arrives at the relation

$$D = D(0) exp[b'\delta_C] = D_0 exp\left(\frac{-Q}{RT}\right) exp\left(\left(\frac{W'}{T}\right)^2 \delta_C\right) \quad (4)$$

or

$$D = D_0 exp\left(\frac{-Q}{RT}\right) exp\left(\left(\frac{W}{T}\right)^2 x_C\right). \quad (4')$$

Here $b'$, $W'$, and $W$ are the corresponding constants. Equations (4) or (4') predict a linear enhancement of the logarithm of the diffusion coefficients of the substitutional atoms with increasing C-induced lattice distortions or C content. In Fig. 5, the measured tracer diffusion coefficients are shown as function of the C concentration in the alloys. The concentration dependencies are plotted in Fig. S2 (Supplementary), too. The dashed lines correspond to predictions according to Eqs. (2)–(4).

For the estimates, we fitted the measured diffusion coefficients using Eq. (4'). An acceptable agreement is seen, Fig. 5 (or Fig. S2). Since migration volume, $\Omega_m$, or the change of the barrier height due to elastic strains, *K*, depend on the type of diffusing atom, a quantitatively different impact of the C atoms on the diffusion of different substitutional atoms is expected. We analysed the measured diffusion coefficients in two way. In Table 3, the diffusion parameters, the pre-exponential factor $D_0$ and the activation energy *Q*, determined as Arrhenius parameters for each alloy are given. Additionally, we have used Eq. (4) and determined the parameters $D_0$, *Q*, and *W* for each diffusing element. The results are listed in Table 4.

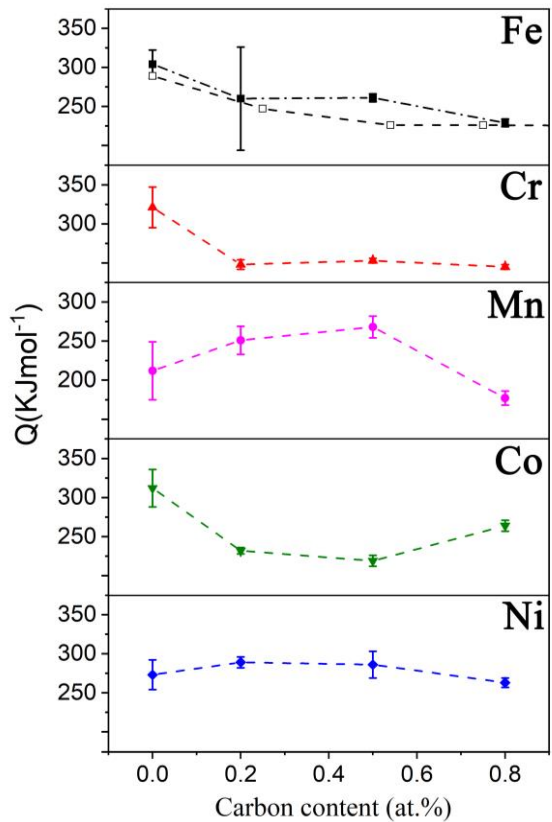

Fig. 6. Effective activation enthalpy of diffusion of $^{59}$Fe (a), $^{51}$Cr (b), $^{63}$Ni (c), $^{57}$Co (d), and $^{54}$Mn (e) in CoCrFeMnNi HEAs as function of the C concentration. In (a), open squares correspond to Fe diffusion in Fe-C alloys [25].

It seems that Eqs. (2)-(4) correctly predict the volume diffusion coefficients for the considered group of alloys, especially at temperatures below 1300 K. The open circles in Fig. 5 correspond to predictions of the fit according to Eq. (4') and the measured diffusivities (filled spheres) almost follow the predictions and increase linearly with increasing C content. The parameter *W* (which is a measure of the impact of interstitial C on migration energies) turned out to be almost identical for all elements, about 12000 K, excepting that for Mn, which features a larger value of about 15000 K. We conclude that diffusion of Mn as an element with the largest radius is most strongly affected by the dissolved C atoms. This reasonable correlation supports indirectly the correctness of the model, Eqs. (2)–(4).

An additional tendency is seen at the highest temperature of 1373 K, where the tracer diffusion coefficients (especially of Ni and Co) decrease first at low C concentrations and start to increase with $x_C \geq 0.5$ at.%, especially for slow-diffusing elements like Co and Ni. We may speculate and associate this behaviour with additional C-induced suppressing of anharmonic effects which in their turn increase the vacancy concentration in C-free alloy at high temperatures. Note that an increase of the vacancy concentration due to temperature-dependent vacancy formation entropy was predicted by careful ab initio calculations for pure Al or Cu [46] or Ni [47]. Thus, at high temperatures, the C alloying is suggested to decrease first the effective vacancy concentration and this trend is reversed at higher C concentrations by the above-discussed impact of interstitial C atoms on the jump barriers.

### 4.1.1 Fe and Cr diffusion

Diffusion of Fe$^{59}$ in the reported materials seems to be extremely interesting. In fact, we guess that one of the closest materials to our alloy is γ-iron from the point of view of the similarity of the crystal lattice and importance of the issue for austenitic steals. It was reported [48, 49] that an increase of the carbon content in iron leads to an increase of the Fe diffusion coefficient of γ-iron, that perfectly correlates with the results presented in our work.

The impact of the carbon content on the activation energies of Fe diffusion in our materials and in γ-iron [25] is compared in Fig. 6 and very similar dependencies are observed.

As a general tendency, a decrease of the effective activation energy of Fe diffusion with an increase of the C content can be identified, Fig. 6. Following the model of Overhauser [26], the lattice distortions induced by the interstitially dissolved C atoms decreases the migration barriers, see Eq. (2). The term $e^{b \cdot x_C}$ contributes to the decrease of the effective activation energy of diffusion by the term $\Delta Q = -k_B \left( \frac{\partial b}{\partial \left(\frac{1}{T}\right)} \right) x_C$. Since the diffusion rates are measured in a relatively narrow temperature interval from 1173 to 1373 K, potential deviations from the linear Arrhenius dependencies are not quantitatively analyzed in the present work. A further analysis would require

Table 3. Arrhenius parameters of volume diffusion (the pre-exponential factor, $D_0$ (in $10^{-4}$ m$^2$/s) and the activation enthalpy, $Q$ (in kJ/mol), of Co, Cr, Fe, Mn and Ni diffusion in CoCrFeMnNi–C HEAs with different content of carbon.

| | Co | Cr | Fe | Mn | Ni |
|---|---|---|---|---|---|
| C = 0 at.% | | | | | |
| $Q$ | 312±24 | 321±26 | 304±18 | 212±37 | 273±19 |
| $D_0$ | $2.64^{+17.60}_{-0.40}$ | $8.50^{+22.40}_{-0.32}$ | $0.46^{+7.77}_{-0.28}$ | $8.01^{+81.30}_{-0.08}$ | $15.60^{+18.50}_{-0.13}$ |
| C = 0.2 at.% | | | | | |
| $Q$ | 232±4 | 248±6 | 260±66 | 251±18 | 289±7 |
| $D_0$ | $0.01^{+0.02}_{-0.01}$ | $0.02^{+0.06}_{-0.01}$ | $0.01^{+0.01}_{-0.01}$ | $0.02^{+0.90}_{-0.01}$ | $5.53^{+7.17}_{-4.26}$ |
| C = 0.5 at.% | | | | | |
| $Q$ | 219±7 | 253±3 | 261±5 | 268±14 | 286±17 |
| $D_0$ | $0.01^{+0.02}_{-0.01}$ | $0.28^{+0.45}_{-0.18}$ | $0.21^{+0.70}_{-0.07}$ | $5.53^{+7.17}_{-4.26}$ | $1.95^{+23.80}_{-0.16}$ |
| C = 0.8 at.% | | | | | |
| $Q$ | 264±7 | 245±3 | 229±5 | 177±9 | 263±6 |
| $D_0$ | $0.35^{+0.77}_{-0.16}$ | $0.16^{+0.24}_{-0.11}$ | $0.01^{+0.02}_{-0.01}$ | $0.01^{+0.01}_{-0.01}$ | $0.24^{+0.49}_{-0.11}$ |

Table 4. Parameters $D_0$ (in $10^{-5}$ m$^2$/s), $Q$ (in kJ/mol) and $W$ (in $10^3$ K) (Eq. (4')) of substitutional element diffusion in the C-alloyed CoCrFeMnNi HEAs.

| | Co | Cr | Fe | Mn | Ni |
|---|---|---|---|---|---|
| $Q$ | 261±12 | 250±13 | 237±11 | 245±17 | 300±20 |
| $D_0$ | $1.20^{+3.89}_{-0.37}$ | $1.05^{+3.76}_{-0.30}$ | $0.27^{+0.79}_{-0.09}$ | $1.03^{+5.36}_{-0.20}$ | $30.97^{+203.02}_{-4.72}$ |
| $W$ | 12.4±5.8 | 12.9±6.0 | 11.9±5.6 | 15.0±7.0 | 12.4±7.4 |

DFT-informed estimates of the factor $b$ in Eqs. (2)–(4).

In the case of Cr, only a minor tendency concerning the decrease of the activation energy with the increase of the C content is seen, Fig. 6. The changes only slightly exceed the experimental accuracy.

Two opposite tendencies of the impact of C alloying on Fe and Cr diffusion are seen in Fig. 5. In fact, the majority of data, below the temperature of 1300 K, do follow the predictions of Eqs. (2)–(4), i.e. the diffusion rates increase with addition of interstitial C atoms, while a decrease of the tracer diffusion coefficients at low C concentrations is obvious at the highest temperature of the present measurements of 1373 K. Whereas the increasing tendency is well described by Eqs. (2) – (4) in term of the impact of interstitial C on the migration barriers, the diffusion retardation at $T = 1373$ K in low-C alloys need a special consideration. As it was explained above, it may be the C alloying-induced suppressing of the anharmonic contributions to the temperature dependent formation entropy of vacancies in these alloys.

### 4.1.2 Ni and Co diffusion

Ni and Co are the slowest diffusing elements in CoCrFeMnNi at elevated temperatures above 1173 K [31] and this tendency is generally confirmed also for the C-alloyed CoCrFeMnNi in the present work. Alloying by C increases generally the diffusion rates of both Ni and Co, Fig. 5, and decreases the effective activation energies, Fig. 6, respectively.

The trend of increasing diffusivity of Ni and Co with increasing C content is more pronounced at lower temperatures, Fig. 5c and d, although for Ni some drop of the corresponding diffusion rate in low-C alloys might be identified at the higher temperatures.

### 4.1.3 Mn diffusion

Manganese is the fastest isotope from all substitutional elements in C-fee CoCrFeMnNi and this tendency holds for all alloys investigated in the present work. Whereas an increase of the amount of carbon to 0.8 at. % enhances the diffusion coefficients with respect to the C-free alloy by a factor of two, or even stronger increase is seen for Mn especially at the lowest temperature in this investigation. We note that the increase of the C amount to 0.8 at.% was found to intensify the grain boundary diffusion contribution, especially for Ni and Co. We are highlighting that the impact of C on the self-diffusion values of the constituent elements is continuous and linear over the significant range of C concentrations, while indicates the high relative stability of the atomic structure of the alloy against carbon alloying.

### 4.2. General tendencies

Summarizing the present findings, several statements are highlighted:

- The volume diffusion coefficients of substitutional elements in the alloy with the highest content of carbon, i.e. in CoCrFeMnNi–0.8 at.% C are slightly larger than those in the C-free alloy by a factor of two to five depending on the temperature and the diffusing element.
- Two tendencies are observed. At lower temperatures, $T < 1300$ K, the diffusion rates of all substitutional elements increase generally with the increasing content of C. On the other hand, a slight decrease of the volume diffusion rates for low-C alloys is found at 1373 K. At this temperature, the slowest diffusion coefficients are recorded for the CoCrFeMnNi–0.2 at.% C alloy.
- The activation energies of diffusion are gradually decreasing with increasing C content. The effect is most pronounced for Mn.

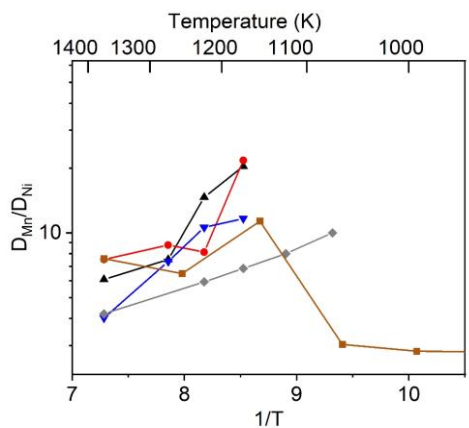

Fig. 7. The radio of the diffusion coefficients of Mn (fastest) and Ni (slowest) in the alloys under investigation: C-free CoCrFeMnNi (grey diamonds [29, 30] and brown squares [31]), and the alloys with 0.2 (black triangles up), 0.5 (red circles) and 0.8 (blue triangles down) at.% of carbon.

In Fig. 7, the ratios of the diffusion coefficients of Mn and Ni in different C-alloyed HEAs are plotted as function of temperature. At elevated temperatures above 1100 K a general trend of increasing ratios of $D_{Mn}/D_{Ni}$ with increasing C content is seen. In the C-free alloy, this trend is drastically reversed at lower temperatures [31] and this behavior has been related to the appearance of a short-range order in the CoCrFeMnNi HEA. In C-alloyed HEAs, the interstitially dissolved C atoms affect differently the jump barriers for Mn and Ni atoms, in general lowering them. Note that both diffusion coefficients, $D_{Mn}$ and $D_{Ni}$, are increasing with the addition of C. The observed behavior correlates with the relative atomic sizes of Mn and Ni atoms and local distortions induced by dissolved C atoms on these substitutional atoms. We may speculate that the C-induced distortions are larger for larger Mn atoms and the Mn diffusion rates are enhanced more significantly in accordance with Eq. (2).

### 5. Conclusions

The present study indicates the existence of two different mechanisms of the influence of C alloying on substitutional diffusion in CoCrFeMnNi-C HEAs. At 1373 K, addition of a small amount of carbon (in the range from 0 to 0.2 at. %) gives rise to diffusion retardation, probably due to decreasing anharmonic contributions to the temperature-dependent vacancy formation entropy. The diffusion retardation is more pronounced for slow diffusing Ni, whereas it is only marginal for fast diffusing elements (e.g., Mn, Cr and Fe). The effective activation energy of diffusion decreases significantly in these alloys and changes only slightly at higher C contents for almost all elements. A continuous decrease of the activation energy of diffusion is observed only for Ni in the whole investigated range of C concentrations.

At lower temperatures a general trend of enhancement of diffusion rates of substitutional elements is clearly seen. This trend is explained by the elastic distortions induced by interstitially dissolved carbon. The enhancement of Mn atoms is largest and this behaviour correlates with largest atomic radius of Mn atoms. The smallest distortions are expected for the solvent atoms with smallest atomic radii, i.e. Co and Ni. Correspondingly, the C-induced enhancements of Co and Ni diffusion in CoCrFeMnNi HEAs are smallest.

The model derived for austenitic steels seems to be applicable for the present interstitially alloyed HEAs and it describes well the change of the crystalline lattice parameter with C alloying.

**Acknowledgments**. O.L. would like to thank Alexander-von-Humboldt Foundation for providing a research fellowship. Financial support from the German Science Foundation (DFG) via research project DI 1419/13-2 is acknowledged. Z. R. would like to thank the China Scholarship Council for providing a PhD scholarship. Z.L. would like to acknowledge the financial support by the National Natural Science Foundation of China (Grant No. 51971248).

# Self-diffusion in carbon-alloyed CoCrFeMnNi high entropy alloys

O.A. Lukianova[1,*], V. Kulitckii[1], Z. Rao[2], Z. Li[2,3], G. Wilde[1], S.V. Divinski[1,**]

[1]Institute of Materials Physics, University of Münster, Germany
[2]Max-Planck-Institut für Eisenforschung, Düsseldorf, Germany
[3]School of Materials Science and Engineering, Central South University, Changsha, China

Here we document the temperature (Fig. S1) and concentration (Fig. S2) dependencies of the diffusion rates of substitutional elements in C-alloyed HEAs.

In Fig. S1, the diffusion rates determined for C-free CoCrFeMnNi alloy in Ref. [31] are shown by brown diamonds. The green squares, black triangles up, red circles and blue triangles down present the results of substitutional element diffusion in the alloys with 0, 0.2, 0.5 and 0.8 at.% of carbon, measured in the present work, respectively.

The solid lines correspond to 'standard' Arrhenius fits of our experimental results using the data for each alloy separately.

Furthermore, we perform a master fit of all data for the given element using Eq. (4) of the main manuscript. The results are presented by green dashed-dot lines (C-free alloy), dotted black lines (alloy with 0.2 at. % of carbon), red dotted lines (alloy with 0.5 at. % of carbon), and dashed blue line (alloy with 0.8 at. % of carbon).

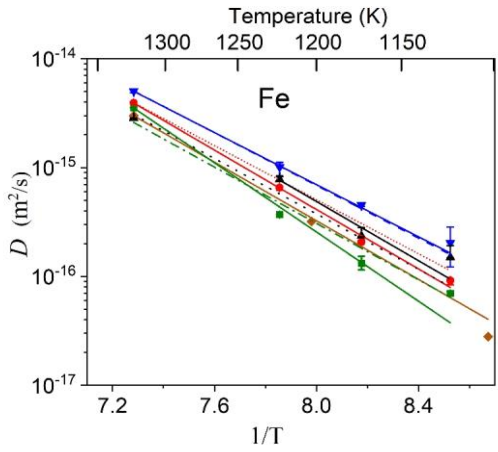

a)

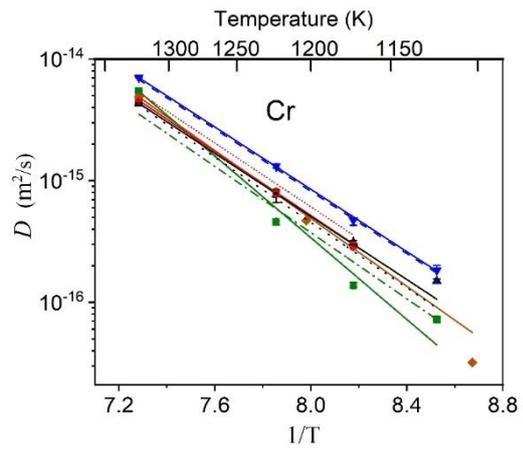

b)

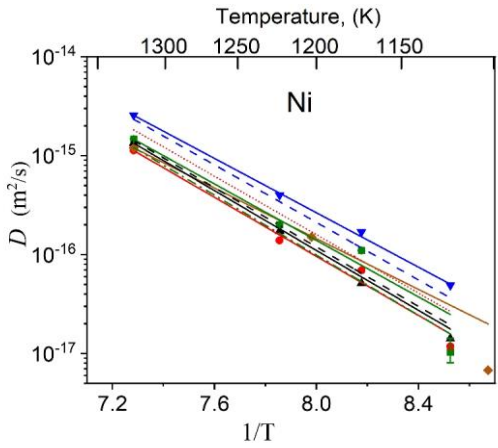

c)

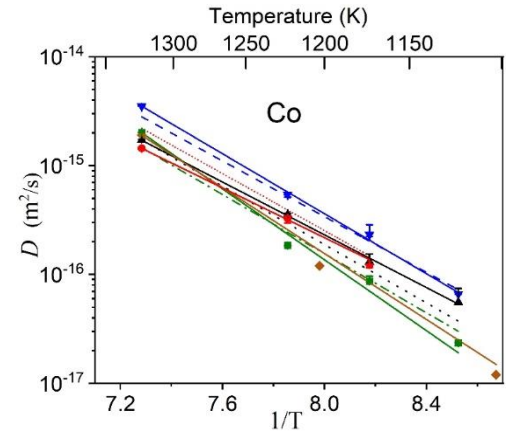

d)

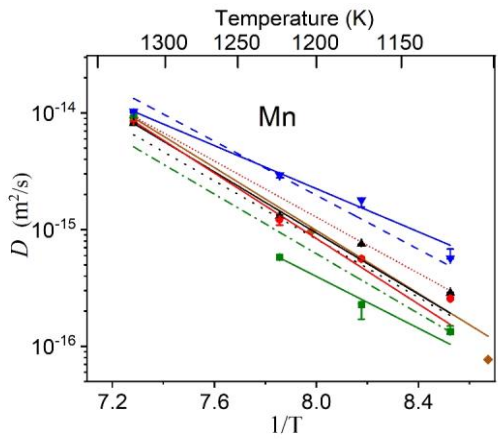

e)

Fig. S1. Arrhenius plots for $^{59}$Fe (a), $^{51}$Cr (b), $^{63}$Ni (c), $^{57}$Co (d), and $^{54}$Mn (e) in polycrystalline CoCrFeMnNi HEAs. The C amount in the alloys are indicated in at.%.

The concentration dependencies of the determined diffusion coefficients are shown in Fig. S2. The solid lines present the anticipated trends and the dashed lines correspond to the master fit using Eq. (4)

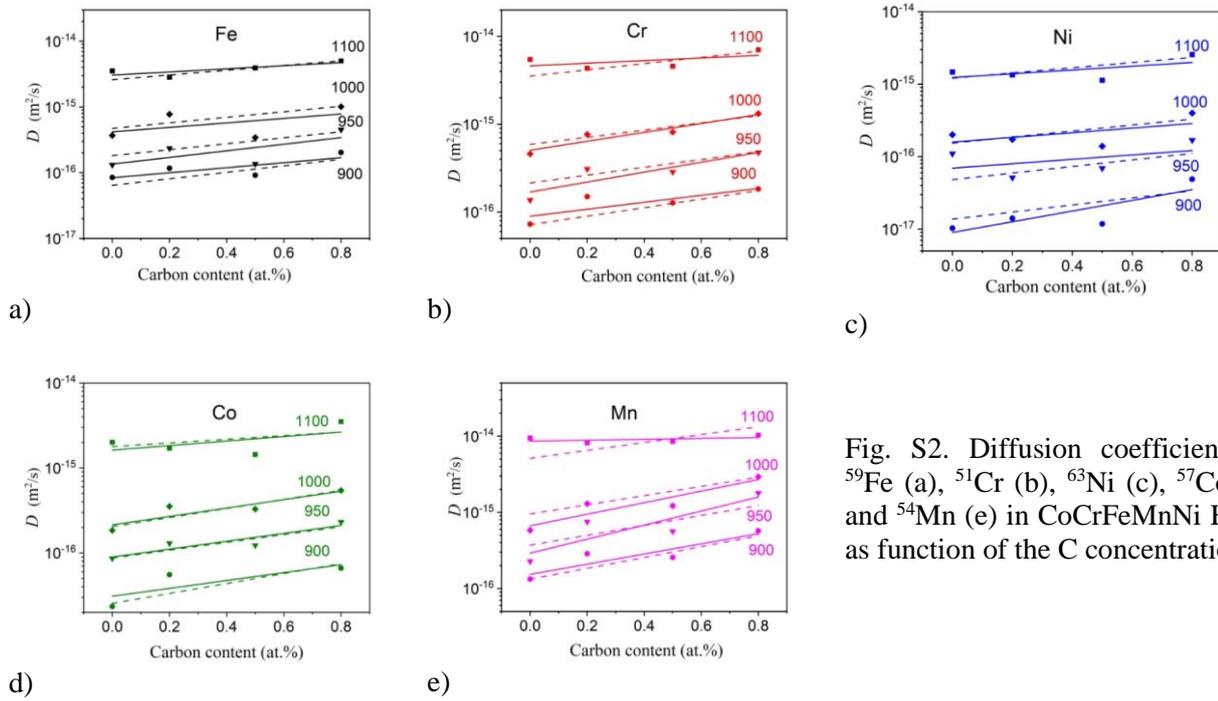

Fig. S2. Diffusion coefficients of $^{59}$Fe (a), $^{51}$Cr (b), $^{63}$Ni (c), $^{57}$Co (d), and $^{54}$Mn (e) in CoCrFeMnNi HEAs as function of the C concentration.